\documentclass[fleqn,usenatbib]{mnras}

\usepackage{newtxtext,newtxmath}

\usepackage[T1]{fontenc}

\DeclareRobustCommand{\VAN}[3]{#2}
\let\VANthebibliography\thebibliography
\def\thebibliography{\DeclareRobustCommand{\VAN}[3]{##3}\VANthebibliography}

\usepackage{graphicx}	
\usepackage{amsmath}	
\usepackage{multirow}
\usepackage{multicol}

\title[E(2)-equivariant radio galaxy classification]{Fanaroff-Riley classification of radio galaxies using group-equivariant convolutional neural networks.}

\author[A.~M.~M.~Scaife \& F.~Porter]{
Anna M. M. Scaife,$^{1,2}$\thanks{E-mail: anna.scaife@manchester.ac.uk (AMS)}
and Fiona Porter$^{1}$
\\
$^{1}$Jodrell Bank Centre for Astrophysics, Department of Physics \& Astronomy, University of Manchester, Oxford Road, Manchester M13 9PL UK \\
$^{2}$ The Alan Turing Institute, Euston Road, London NW1 2DB, UK 
}

\date{Accepted XXX. Received YYY; in original form ZZZ}

\pubyear{2015}

\begin{document}
\label{firstpage}
\pagerange{\pageref{firstpage}--\pageref{lastpage}}
\maketitle

\begin{abstract}
Weight sharing in convolutional neural networks (CNNs) ensures that their feature maps will be translation-equivariant. However, although conventional convolutions are equivariant to translation, they are not equivariant to other isometries of the input image data, such as rotation and reflection. For the classification of astronomical objects such as radio galaxies, which are expected statistically to be globally orientation invariant, this lack of dihedral equivariance means that a conventional CNN must learn explicitly to classify all rotated versions of a particular type of object individually. In this work we present the first application of group-equivariant convolutional neural networks to radio galaxy classification and explore their potential for reducing intra-class variability by preserving equivariance for the Euclidean group E(2), containing translations, rotations and reflections. For the radio galaxy classification problem considered here, we find that classification performance is modestly improved by the use of both cyclic and dihedral models without additional hyper-parameter tuning, and that a $D_{16}$ equivariant model provides the best test performance. We use the Monte Carlo Dropout method as a Bayesian approximation to recover  epistemic uncertainty as a function of image orientation and show that E(2)-equivariant models are able to reduce variations in model confidence as a function of rotation. 
\end{abstract}




\section{Introduction}

In radio astronomy, a massive increase in data volume is currently driving the increased adoption of machine learning methodologies and automation during data processing and analysis. This is largely due to the high data rates being generated by new facilities such as the Low-Frequency Array  \citep[LOFAR;][]{VanHaarlem2013}, the Murchison Widefield Array \citep[MWA;][]{Beardsley2019}, the MeerKAT telescope \citep{Jarvis2016}, and the Australian SKA Pathfinder (ASKAP) telescope \citep{Johnston2008}. For these instruments a natural solution has been to automate the data processing stages as much as possible, including classification of sources.

With the advent of such huge surveys, new automated classification algorithms have been developed to replace the \emph{``by eye''} classification methods used in earlier work. In radio astronomy, morphological classification using convolutional neural networks (CNNs) and deep learning is becoming increasingly common for object classification, in particular with respect to the classification of radio galaxies. The ground work in this field was done by \cite{aniyan2017} who made use of CNNs for the classification of Fanaroff-Riley (FR) type I and type II radio galaxies \citep{fr1974}. This was followed by other works involving the use of deep learning in source classification. Examples include  \cite{lukic2018} who made use of CNNs for the classification of compact and extended radio sources from the Radio Galaxy Zoo catalogue \citep{rgz}, the CLARAN \citep[Classifying Radio Sources Automatically with a Neural Network;][]{wu2018} model made use of the Faster R-CNN \citep{fasterrcnn} network to identify and classify radio sources; \cite{alger2018} made use of an ensemble of classifiers including CNNs to perform host galaxy cross-identification. \cite{hmtnet} made use of transfer learning with CNNs to perform cross-survey classification, while \cite{gheller2018} made use of deep learning for the detection of cosmological diffuse radio sources. \cite{lukic2018} also performed morphological classification using a novel technique known as capsule networks \citep{sabour2017}, although they found no specific advantage compared to traditional CNNs. \cite{bowles2020} showed that an attention-gated CNN could be used to perform Fanaroff-Riley classification of radio galaxies with equivalent performance to other applications in the literature, but using $\sim$50\% fewer learnable parameters than the next smallest classical CNN in the field. 

Convolutional neural networks classify images by learning the weights of convolutional kernels via a training process and using those learned kernels to extract a hierarchical set of feature maps from input data samples. Convolutional weight sharing makes CNNs more efficient than multi-layer perceptrons (MLPs) as it ensures translation-equivariant feature extraction, i.e. a translated input signal results in a corresponding translation of the feature maps. However, although conventional convolutions are equivariant to translation, they are not equivariant to other isometries of the input data, such as rotation, i.e. rotating an image and then convolving with a fixed filter is not the same as first convolving and then rotating the result. Although many CNN training implementations use rotation as a form of data augmentation, this lack of rotational equivariance means that a conventional CNN must explicitly learn to classify all rotational augmentations of each image individually. This can result in CNNs learning multiple copies of the same kernel but in different orientations, an effect that is particularly notable when the data itself possesses rotational symmetry \citep{dieleman2016}. Furthermore, while data augmentation that mimicks a form of equivariance, such as image rotation, can result in a network learning approximate equivariance if it has sufficient capacity, it is not guaranteed that invariance learned on a training set will generalise equally well to a test set \citep{lencvedaldi}. A variety of different equivariant networks have been developed to address this issue, each guaranteeing a particular transformation equivariance between the input data and associated feature maps. For example, in the field of galaxy classification using optical data, \cite{dieleman2015} enforced discrete rotational invariance through the use of a multi-branch network that concatenated the output features from multiple convolutional branches, each using a rotated version of the same data sample as its input. However, while effective, the approach of \cite{dieleman2015} requires the convolutional layers of a network architecture and hence the number of model weights associated with them to be replicated $N$ times, where $N$ is the number of discrete rotations.

Recently, a more efficient method of using convolutional layers that are equivariant to a particular group of transforms has been developed, which requires no replication of architecture and hence fewer learnable parameters to be used. Explicitly enforcing an equivariance in the network model in this way not only provides a guarantee that it will generalise, but also prevents the network using parameter capacity to learn characteristic behaviour that can instead be specified a priori. First introduced by \cite{cohenwelling2016}, these Group equivariant Convolutional Neural Networks (G-CNNs), which preserve group equivariance through their convolutional layers, are a natural extension of conventional CNNs that ensure translational invariance through weight sharing. Group equivariance has also been demonstrated to improve generalisation and increase performance \citep[see e.g.][]{weiler2017, weilercesa2019}. In particular, \emph{Steerable} G-CNNs have become an increasingly important solution to this problem and notably those steerable CNNs that describe E(2)-equivariant convolutions. 

The Euclidean group E(2) is the group of isometries of the plane $\mathbb{R}^2$ that contains translations, rotations and reflections. Isometries such as these are important for general image classification using convolution as the target object in question is unlikely to appear at a fixed position and orientation in every test image. Such variations are not only highly significant for objects/images that have a preferred orientation, such as text or faces, but are also important for low-level features in nominally orientation-unbiased targets such as astrophysical objects. In principle, E(2)-equivariant CNNs will generalize over rotationally-transformed images by design, which reduces the amount of intra-class variability that they have to learn. In effect such networks are insensitive to rotational or reflection variations and therefore learn only features that are independent of these properties.

In this work we introduce the use of $G$-steerable CNNs to astronomical classification. The structure of the paper is as follows: in Section~\ref{sec:gsteer} we describe the mathematical operation of $G$-steerable CNNs and define the specific Euclidean subgroups being considered in this work; in Section~\ref{sec:data} we describe the data sets used in this work and the preprocessing steps implemented on those data; in Section~\ref{sec:arch} we describe the network architecture adopted in this work, explain how the $G$-steerable implementation is constructed and specify the group representations; in Section~\ref{sec:results} we give an overview of the training outcomes including a discussion of the convergence for different equivalence groups, validation and test performance metrics, and introduce a novel use of the Monte Carlo Dropout method for quantitatively assessing the degree of model confidence in a test prediction as a function of image orientation; in Section~\ref{sec:discussion} we discuss the validity of the assumptions that radio galaxy populations are expected to be staitsically rotation and reflection unbiased and review the implications of this work in that context; in Section~\ref{sec:conc} we draw our conclusions.

\section{E(2)-Equivariant G-Steerable CNNs}
\label{sec:gsteer}

Group CNNs define feature spaces using feature fields $f : \mathbb{R}^2 \rightarrow \mathbb{R}^c$, which associate a $c$-dimensional feature vector $f(x) \in \mathbb{R}^c$ to each point $x$ of an input space. Unlike conventional CNNs, the feature fields of such networks contain transformations that preserve the transformation law of a particular group or subgroup, which allows them to encode orientation information. This means that if one transforms the input data, $x$, by some transformation action, $g$, (translation, rotation, etc.) and passes it through a trained layer of the network, then the output from that layer, $\Phi(x)$, must be equivalent to having passed the data through the layer and then transformed it, i.e.
\begin{equation}
\label{eq:equivariance}
    \Phi(\mathcal{T}_g x) = \mathcal{T}'_g \Phi(x),
\end{equation}
where $\mathcal{T}_g$ is the transformation for action $g$. In the case where the transformation is \emph{invariant} rather than \emph{equivariant}, i.e. the input does not change at all when it is transformed, $\mathcal{T}'_g$ will be the identity matrix for all actions $g \in G$. In the case of equivariance, $\mathcal{T}_g$ does not necessarily need to be equal to $\mathcal{T}'_g$ and instead must only fulfil the property that it is a linear representation of $G$, i.e. $\mathcal{T}(gh) = \mathcal{T}(g)\mathcal{T}(h)$.

\cite{cohenwelling2016} demonstrated that the conventional convolution operation in a network can be re-written as a group convolution:
\begin{equation}
    [f \ast \phi](g) = \sum_{h \in X} \sum_k {f_k(h) \phi_k(g^{-1}h)},
\end{equation}
where $X = \mathbb{R}^2$ in layer one and $X = G$ in all subsequent layers. Whilst this operation is translationally-equivariant, $\phi$ is still rotationally constrained. For E(2)-equivariance to hold more generally, the kernel itself must satisfy 
\begin{equation}
    \phi(gx) = \rho_{\rm out}(g) \phi(x) \rho_{\rm in}(g^{-1})~~~\forall\,g \in G,~x \in \mathbb{R}^2, 
\end{equation}
\citep{weiler2018}, where $g$ is an action from group $G$, and $\phi : \mathbb{R}^2 \rightarrow \mathbb{R}^{c_{\rm in} \times c_{\rm out}}$, where $c_{\rm in}$ and $c_{\rm out}$ are the number of channels in the input and output data, respectively; $\rho$ is the group representation, which specifies how the channels of each feature vector mix under transformations. Kernels which fulfil this constraint are known as \emph{rotation-steerable} and must be constructed from a suitable family of basis functions. As noted above, this is a linear relationship, which means that G-steerable kernels form a subspace of the convolution kernels used by conventional CNNs. 

For planar images the input space will be $\mathbb{R}^2$, and for single frequency or continuum radio images these feature fields will be scalar, such that $s : \mathbb{R}^2 \rightarrow \mathbb{R}$. The group representation for scalar fields is also known as the \emph{trivial} representation, $\rho(g) = 1~~\forall~g \in G$, indicating that under a transformation there is no orientation information to preserve and that the amplitude does not change. The group representation of the output space from a G-steerable convolution must be chosen by the user when designing their network architecture and can be thought of as a variety of hyper-parameter.  

However, whilst the representation of the input data is in some senses quite trivial for radio images, in practice convolution layers are interleaved with other operations that are sensitive to specific choices of representation. In particular, the range of non-linear activation layers permissible for a particular group or subgroup representation may be limited. Trivial representations, such as scalar fields, do not transform under rotation and therefore conventional nonlinearities like the widely used ReLU activation function are fine. Bias terms in convolution allow equivariance for group convolutions only in the case where there is a single bias parameter per group feature map (rather than per channel feature map) and likewise for batch normalisation \citep{cohenwelling2016}.

In this work we use the G-steerable network layers from \cite{weilercesa2019} who define the Euclidean group as being constructed from the translation group, $(\mathbb{R}, +)$, and the orthogonal group, O(2)~$= \{ O \in \mathbb{R}^{2\times 2}~|~O^TO = {\rm id}_{2\times2} \}$, such that the Euclidean group is congruent with the semi-direct product of these two groups, E(2)~$ \cong (\mathbb{R}, +) \rtimes$ O(2). Consequently, the operations contained in the orthogonal group are those which leave the origin invariant, i.e. continuous rotations and reflections. In this work we specifically consider the cyclic subgroups of the Euclidean group with form $(\mathbb{R}^2, +) \rtimes C_N$, where $C_N$ contains a set of discrete rotations in multiples of $2\pi/N$, and the dihedral subgroups with form $(\mathbb{R}^2, +) \rtimes D_N$, where $D_N \cong C_N \rtimes (\{\pm 1\}, \ast)$, which incorporate reflection around $x=0$ in addition to discrete rotation. As noted by \cite{cohenwelling2016}, although convolution on continuous groups is mathematically well-defined, it is difficult to approximate numerically in a fully equivariant manner. Furthermore, the complete description of all transformations in larger groups is not always feasible \citep{gensdomingos}. Consequently, in this work we consider only the discrete and comparatively small groups, $C_N$ and $D_N$, with orders $N$ and $2N$, respectively.
\begin{figure*}
\centerline{\includegraphics[width=0.9\textwidth]{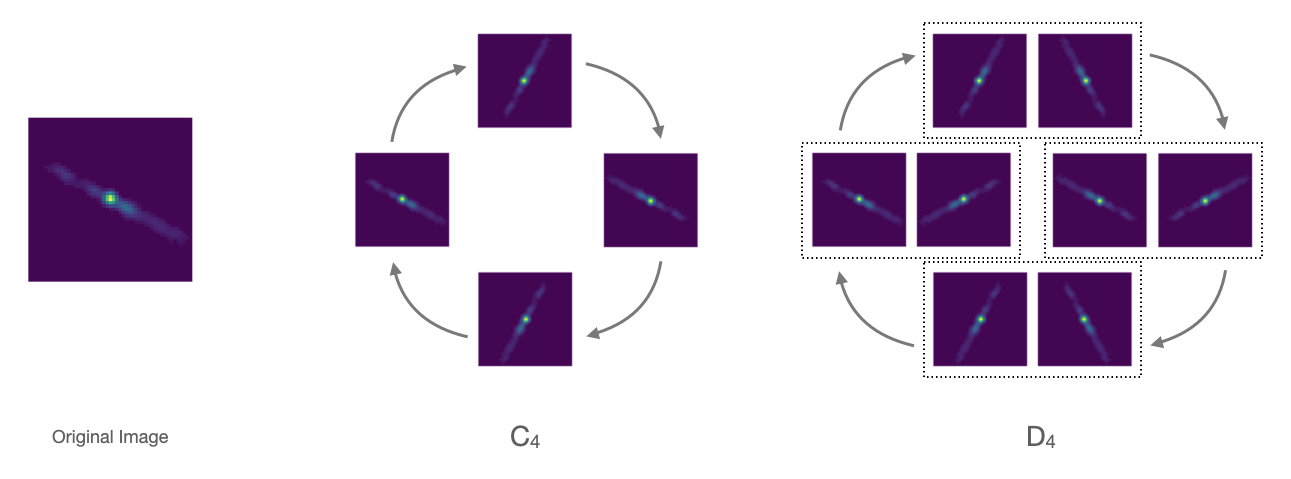}}
\caption{Illustration of the $C_4$ and $D_4$ groups for an example radio galaxy postage stamp image with $50\times 50$\,pixels. The members of the $C_4$ group are each rotated by $\pi/2$\,radians, resulting in a group order $|C_4| = 4$. The members of the $D_4$ group are each rotated by $\pi/2$\,radians and mirrored around $x=0$, resulting in a group order $|D_4| = 8$. \label{fig:finitegroups}}
\end{figure*}

\section{Data}
\label{sec:data}

The data set used in this work is based on the catalogue of \cite{mirabest2017}, who used a parent galaxy sample taken from \cite{bestheckman} that cross-matched the Sloan Digital Sky Survey \citep[SDSS;][]{sdss} data release 7 \citep[DR7;][]{sdssdr7} with the Northern VLA Sky Survey \citep[NVSS;][]{NVSS} and the Faint Images of the Radio Sky at Twenty centimetres \citep[FIRST;][]{FIRST}.

From the parent sample, sources were visually classified by \cite{mirabest2017} using the original morphological definition provided by \cite{fr1974}: galaxies which had their most luminous regions separated by less than half of the radio source’s extent were classed as FRI, and those which were separated by more than half of this were classed as FRII. Where the determination of this separation was complicated by either the limited resolution of the FIRST survey or by its poor sensitivity to low surface brightness emission, the human subjectivity in this calculation was indicated by the source classification being denoted as ``Uncertain", rather than ``Confident". Galaxies were then further classified into morphological sub-types via visual inspection. Any sources which showed FRI-like behaviour on one half of the source and FRII-like behaviour on the other were deemed to be hybrid sources.

Each object within the catalogue of \cite{mirabest2017} was given a three-digit classification identifier to allow images to be separated into different subsets. Images were classified by FR class, confidence of classification, and morphological sub-type. These are summarised in Table~\ref{tab:mbdata}. For example, a radio galaxy that was confidently classified as an FRI type source with a wide-angle tail morphology would be denoted 102.
\begin{table}
    \centering
    \begin{tabular}{lll}
        \hline
         Digit 1 & Digit 2 & Digit 3  \\\hline
         0 - FRI    & 0 - Confident   & 0 - Standard \\
         1 - FRII   & 1 - Uncertain   & 1 - Double-double \\
         2 - Hybrid &                 & 2 - Wide-angle Tail \\
         3 - Unclassifiable &         & 3 - Diffuse \\
                    &                 & 4 - Head-tail \\\hline
         
    \end{tabular}
    \caption{Numerical identifiers from the catalogue of \protect\cite{mirabest2017}. \label{tab:mbdata}}
\end{table}

We note that not all combinations of the three digits described in Table~\ref{tab:mbdata} are present in the catalogue as some morphological classes are dependent on the parent FR class, with only FRI type objects being sub-classified into head-tail or wide-angle tail, and only FRII type objects being sub-classified as double-double. Hybrid FR sources are not considered to have any non-standard morphologies, as their standard morphology is inherently inconsistent between sources. Confidently classified objects outnumber their uncertain counterparts across all classes, and in classes that have few examples there may be no uncertain sources present. This is particularly apparent for non-standard morphologies.

From the full catalog of 1329 labelled objects, 73 were excluded from the machine learning data set. These include (i) the 40 objects denoted as \emph{3 - unclassifiable}, (ii) 28 objects which had an angular extent greater than a selected image size of $150\times150$ pixels, (iii) 4 objects with structure that was found to overlap the edge of the sky area covered by the FIRST survey, and (iv) the single object in 3-digit category 103. This final object was excluded as a minimum of two examples from each class are required for the data set: one for the training set and one for the test set. Following these exclusions, 1256 objects remain, which we refer to as the \emph{MiraBest} data set and summarise in Table~\ref{tab:mb}.
\begin{table*}
    \centering
    \begin{tabular}{|c|c|c|c|c|c|c|}
        \hline
         Class & No. &  & Confidence & Morphology & No. & MiraBest Label \\ 
        \hline\hline
        \multirow{5}{*}{FRI} & \multirow{5}{*}{591} & \multirow{5}{*}{} & \multirow{3}{*}{Confident} & Standard & 339 & 0 \\ \cline{5-7}
         &  &  &  & Wide-Angle Tailed & 49 & 1 \\ \cline{5-7}
         &  &  &  & Head-Tail & 9 & 2 \\ \cline{4-7}
         &  &  & \multirow{2}{*}{Uncertain} & Standard & 191 & 3 \\ \cline{5-7}
         &  &  &  & Wide-Angle Tailed & 3 & 4 \\  \hline
        \multirow{3}{*}{FRII} & \multirow{3}{*}{631} & \multirow{3}{*}{} & \multirow{2}{*}{Confident} & Standard & 432 & 5 \\ \cline{5-7}
         &  &  &  & Double-Double & 4 & 6 \\ \cline{4-7}
         &  &  & Uncertain & Standard & 195 & 7 \\  \hline
        \multirow{2}{*}{Hybrid} & \multirow{2}{*}{34} & \multirow{2}{*}{} & Confident & NA & 19 & 8 \\ \cline{4-7}
         &  &  & Uncertain & NA & 15 & 9 \\ \hline
    \end{tabular}
\caption{MiraBest data set summary. The original data set labels (MiraBest Label) are shown in relation to the labels used in this work (Label). Hybrid sources are not included in this work, and therefore have no label assigned to them. \label{tab:mb}}
\end{table*}

All images in the \emph{MiraBest} data set are subjected to a similar data pre-processing as other radio galaxy deep learning data sets in the literature \citep[see e.g.][]{aniyan2017, hmtnet}. FITS images for each object are extracted from the FIRST survey data using the Skyview service \citep{skyview} and the {\tt astroquery} library \citep{astroquery}. These images are then processed in four stages before data augmentation is applied: firstly, image pixel values are set to zero if their value is below a threshold of three times the local rms noise, secondly the image size is clipped to 150 by 150 pixels, i.\,e. $270^{\prime\prime}$ by $270^{\prime\prime}$ for FIRST, where each pixel corresponds to $1.8^{\prime\prime}$. Thirdly, all pixels outside a square central region with extent equal to the largest angular size of the radio galaxy are set to zero. This helps to eliminate secondary background sources in the field and is possible for the \emph{MiraBest} data set due to the inclusion of this parameter in the catalogue of \cite{mirabest2017}. Finally the image is normalised as:
\begin{equation}
    \text{Output} = 255\cdot\frac{\text{Input} - \text{min}(\text{Input})}{\text{max}(\text{Input})-\text{min}({\text{Input}})},
\label{eqn:pre_processing_norm}
\end{equation}
where `Output' is the normalised image, `Input' is the original image and `min' and `max' are functions which return the single minimal and maximal values of their inputs, respectively. Images are saved to PNG format and accummulated into a {\tt PyTorch} batched data set\footnote{The MiraBest data set is available on Zenodo: 10.5281/zenodo.4288837}.

For this work we extract the objects labelled as Fanaroff-Riley Class~I (FRI) and Fanaroff-Riley Class~II \citep[FRII;][]{fr1974} radio galaxies with classifications denoted as Confident (as opposed to Uncertain). We exclude the objects classified as Hybrid and do not employ sub-classifications. This creates a binary classification data set with target classes FRI and FRII. 
We denote the subset of the full \emph{MiraBest} data set used in this work as \emph{MiraBest$^{\ast}$}. 

The \emph{MiraBest$^{\ast}$} data set has pre-specified training and test data partitions and the number of objects in each of these partitions is shown in Table~\ref{tab:data} along with the equivalent partitions for the full \emph{MiraBest} data set. In this work we subdivide the \emph{MiraBest*} training partition into training and validation sets using an 80:20 split. The test partition is reserved for deriving the performance metrics presented in Section~\ref{sec:performance}.

To accelerate convergence, we further normalise individual data samples from the data set by shifting and scaling as a function of the mean and variance, both calculated from the full training set \citep{tricks} and listed in Table~\ref{tab:data}.
\begin{table}
    \centering
    \caption{Data used in this work. The table shows the number of objects of each class that are provided in the training and test partitions for the \emph{MiraBest} data set, containing sources labeled as both Confident and Uncertain, and the \emph{MiraBest$^{\ast}$} data set, containing only objects labeled as Confident, as well as the mean and standard deviation of the training sets in each case.  \label{tab:data}}
    \begin{tabular}{@{\extracolsep{3pt}}lcccccc@{}}
    \hline
    & \multicolumn{2}{c}{Train} & \multicolumn{2}{c}{Test} & & \\
    \cline{2-3} \cline{4-5}
    Data & FRI & FRII & FRI & FRII & $\mu$ & $\sigma$ \\\hline
    \emph{MiraBest} & 517 & 552 & 74 & 79 & 0.0031 & 0.0352 \\
    \emph{MiraBest$^{\ast}$}   & 348 & 381 & 49 & 55 & 0.0031 & 0.0350 \\\hline
    \end{tabular}
\end{table}
Data augmentation is performed during training and validation for all models using random rotations from 0 to 360\,degrees. This is standard practice for augmentation and is also consistent with the $G$-steerable CNN training implementations of \cite{weilercesa2019}, who included rotational augmentation for their own tests in order to not disadvantage models with lower levels of equivariance. To avoid issues arising from samples where the structure of the radio source overlaps the edge of the field and is artificially truncated in some orientations during augmentation, but not in others, we apply a circular mask to each sample image, setting all pixels to zero outside a radial distance from the centre of 75\,pixels.

An example data sample is shown in Figure~\ref{fig:finitegroups}, where it is used to illustrate the corresponding $C_4$ and $D_4$ groups. As noted by \cite{weilercesa2019}, for signals digitised on a pixel grid, exact equivariance is not possible for groups that are not symmetries of the grid itself and in this case only subgroups of $D_4$ will be exact symmetries with all other subgroups requiring interpolation to be employed \citep{dieleman2016}.

\section{Architecture}
\label{sec:arch}

For our architecture we use a simple LeNet-style network \citep{lenet} with two convolutional layers, followed by three fully-connected layers. Each of the convolutional layers has a ReLU activation function and is followed by a max-pooling operation. The fully-connected layers are followed by ReLU activation functions and we use a 50\% dropout before the final fully-connected layer, as is standard for LeNet \citep{lenetdropout}. An overview of the  architecture is shown in Table~\ref{tab:lenet}. In what follows we refer to this base architecture using conventional convolution layers as the \emph{standard CNN} and denote it $\{e\}$. We also note that the use of \emph{conventional CNN} is used through the paper to refer to networks that do not employ group-equivariant convolutions, independent of architecture.

For the $G$-steerable implementation of this network we use the {\tt e2cnn} extension\footnote{\url{https://github.com/QUVA-Lab/e2cnn}} to the PyTorch library \citep{weilercesa2019} and replace the convolutional layers with their subgroup-equivariant equivalent. We also introduce two additional steps into the network in order to recast the feature data from the convolutional layers into a format suitable for the conventional fully-connected layers. These steps consist of reprojecting the feature data from a geometric tensor into standard tensor format and pooling over the group features, and are indicated in italics in Table~\ref{tab:lenet}. Since the additional steps in the $G$-steerable implementations have no learnable parameters associated with them, the overall architecture is unchanged from that of the standard CNN; it is only the nature of the kernels in the convolutional layers that differ.
\begin{table}
    \centering
    \caption{The LeNet5-style network architecture used for all the models in this work. $G$-Steerable implementations include the additional steps indicated in italics and replace the convolutional layers with the appropriate group-equivariant equivalent in each case. Column [1] lists the operation of each layer in the network, where entries in italics denote operations that are applied only in the $G$-steerable version of the network; Column [2] lists the kernel size in pixels for each layer, where appropriate; Column [3] lists the number of output channels from each layer; Column [4] denotes the degree of zero-padding in pixels added to each edge of an image, where appropriate. \label{tab:lenet}}
    \begin{tabular}{lccc}
    \hline
    Operation & Kernel & Channels & Padding \\\hline
        \emph{Invariant Projection} & & & \\
        Convolution & $5\times5$ & 6 & 1  \\
        ReLU & & & \\
        Max-pool & $2\times2$ & & \\
        Convolution & $5\times5$ & 16 & 1  \\
        ReLU & & & \\
        Max-pool & $2\times2$ & & \\
        \emph{Invariant Projection} & & & \\
        \emph{Global Average Pool} & & & \\
        Fully-connected & & 120 & \\
        ReLU & & & \\
        Fully-connected & & 84 & \\
        ReLU & & & \\
        Dropout ($p=0.5$)& & & \\
        Fully-connected & & 2 & \\\hline
    \end{tabular}
\end{table}

For the input data we use the trivial representation, but for all subsequent steps in the $G$-steerable implementations we adopt the \emph{regular} representation, $\rho_{\rm reg}$. This representation  is typical for describing finite groups/subgroups such as $C_N$ and $D_N$. The regular representation of a finite group $G$ acts on a vector space $\mathbb{R}^{|G|}$ by permuting its axes, where $|G| = N$ for $C_N$ and $|G| = 2N$ for $D_N$, see Figure~\ref{fig:finitegroups}. This representation is helpful because its action simply permutes channels of fields and is therefore equivariant under pointwise operations such as the ReLU activation function, max and average pooling functions \citep{weilercesa2019}. 

We train each network over 600 epochs using a standard cross-entropy loss function and the Adam optimiser \citep{adam} with an initial learning rate of $10^{-4}$ and a weight decay of $10^{-6}$. We use a scheduler to reduce the learning rate by 10\% each time the validation loss fails to decrease for two consecutive epochs. We use mini-batching with a batch size of 50. No additional hyper-parameter tuning is performed. We also implement an early-stopping criterion based on validation accuracy and for each training run we save the model corresponding to this criterion.

\section{Results}
\label{sec:results}

\begin{figure*}
\centerline{\includegraphics[width=0.5\textwidth]{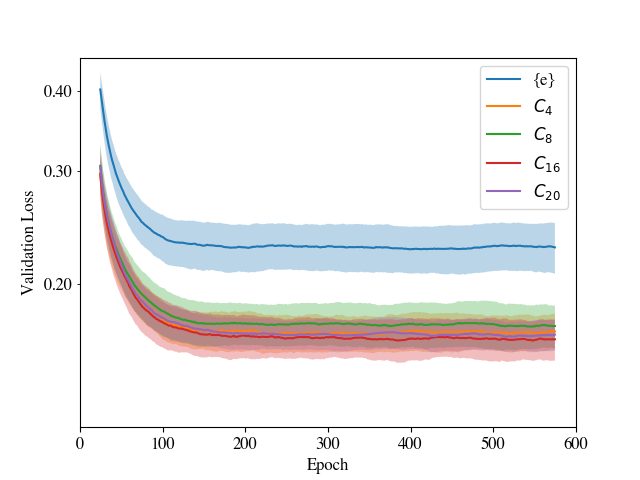}\qquad\includegraphics[width=0.5\textwidth]{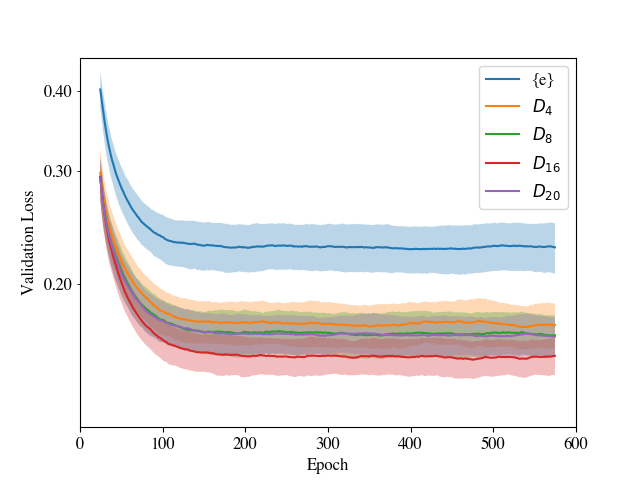}}
\caption{Validation losses during the training of the standard CNN, denoted $\{e\}$, and (i) $C_N$-equivariant models for the MiraBest$^{\ast}$ data set (left), and (ii) $D_N$-equivariant models for the MiraBest$^{\ast}$ data set (right). Plots show mean and standard deviation over five training repeats. Curves are smoothed over 20 epochs to eliminate small-scale variability. \label{fig:loss}}
\end{figure*}

\subsection{Convergence of G-Steerable CNNs}
\label{sec:convergence}

Validation loss curves for both the standard CNN implementation, denoted $\{e\}$, and the group-equivariant CNN implementations for $N = \{4,8,16,20\}$ are shown in Figure~\ref{fig:loss}. Curves show the mean and standard deviation for each network over five training repeats. It can seen from Figure~\ref{fig:loss} that the standard CNN implementation achieves a significantly poorer loss than that of its group-equivariant equivalents. For both the cyclic and dihedral group-equivariant models, the best validation loss is achieved for $N=16$. Although the final loss in the case of the cyclic and dihedral-equivariant networks is not significantly different in value, it is notable that the lower order dihedral networks converge towards this value more rapidly than the equivalent order cyclic networks. We observe that higher order groups minimize the validation loss more rapidly, i.e. the initial gradient of the loss as a function of epoch is steeper, up to order $N=16$ in this case. \cite{weilercesa2019}, who also noted the same thing when training on the MNIST datasets, attribute this behaviour to the increased generalisation capacity of equivariant networks, since there is no significant difference in the number of learnable parameters between models.

\begin{figure}
\centerline{\includegraphics[width=0.5\textwidth]{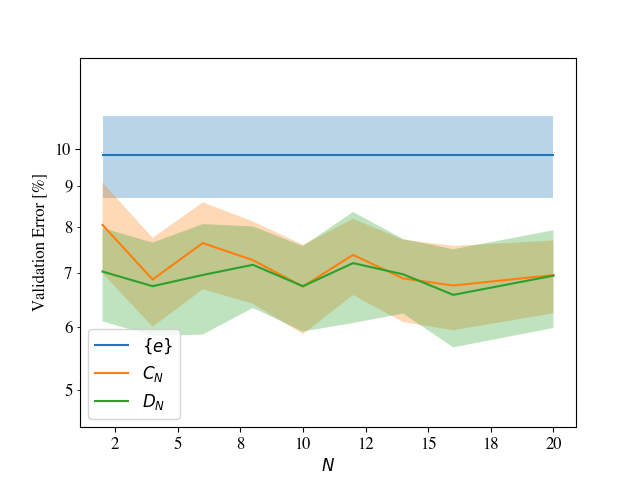}}
\caption{Validation errors of $C_N$ and $D_N$ regular steerable CNNs for different orders, $N$, for the MiraBest$^{\ast}$ data set. All equivariant models improve upon the non-equivariant CNN baseline, $\{e\}$.  \label{fig:err_order}}
\end{figure}

Final validation error as a function of order, $N$, for the group-equivariant networks is shown in Figure~\ref{fig:err_order}. From this figure it can be seen that all equivariant models improve upon the non-equivariant CNN baseline, $\{e\}$, and that the validation error decreases before reaching a minimum for both cyclic and dihedral models at approximately 16 orientations. This behaviour is discussed further in Section~\ref{sec:hp}.

\subsection{Performance of G-Steerable CNNs}
\label{sec:performance}

Standard performance metrics for both the standard CNN implementation, denoted $\{e\}$, and the group-equivariant CNN implementations for $N = \{4,8,16,20\}$ are shown in Table~\ref{tab:frf_metrics}. The metrics in this table are evaluated using the reserved test set of the \emph{MiraBest$^{\ast}$} data set, classified using the best-performing model according to the validation early-stopping criterion. The reserved test set is augmented by a factor of 9 using discrete rotations of 20$^{\circ}$ over the interval $[0^{\circ}, 180^{\circ})$. This augmentation is performed in order to provide metrics that reflect the performance over a consistent range of orientations. The values in the table show the mean and standard deviation for each metric over five training repeats. All $G$-steerable CNNs listed in this table use a regular representation for feature data and apply a $G$-invariant map after the convolutional layers to guarantee an invariant prediction. 
\begin{table*}
    \centering
    \caption{Performance metrics for classification of the MiraBest$^{\ast}$ data set using the standard CNN ($\{e\}$) and $G$-steerable CNNs for different cyclic and dihedral subgroups of the E(2) Euclidean group. All $G$-steerable CNNs use a regular representation for feature data and apply a G-invariant map after the convolutions to guarantee an invariant prediction. \label{tab:frf_metrics}}
   
    \begin{tabular}{@{\extracolsep{3pt}}lccccccc@{}}
    \hline
    &&\multicolumn{3}{c}{FRI} & \multicolumn{3}{c}{FRII} \\
    \cline{3-5} \cline{6-8}
    \textbf{MiraBest$^{\ast}$} & \textbf{Accuracy [\%]} & Precision & Recall & F1-score & Precision & Recall & F1-score \\\hline
    $\{e\}$ & $94.04 \pm 1.37$ &  $0.935 \pm 0.018$ & $0.940 \pm 0.024$ & $0.937 \pm 0.015$ &  $0.946 \pm 0.020$ & $0.941 \pm 0.018$ & $0.944 \pm 0.013$\\\hline
    $C_4$  &  $95.24 \pm 1.23$ &  $0.942 \pm 0.018$ & $0.959 \pm 0.015$ & $0.950 \pm 0.013$ & $0.963 \pm 0.013$ & $0.947 \pm 0.018$ & $0.955 \pm 0.012$\\
    $C_8$  &   $95.96 \pm 1.06$ & $0.950 \pm 0.020$ & $0.966 \pm 0.016$ &  $0.958 \pm 0.011$ & $0.969 \pm 0.013$ & $0.954 \pm 0.019$ & $0.961 \pm 0.010$\\
    $C_{16}$ & $96.07 \pm 1.03$ & $0.953 \pm 0.020$ & $0.964 \pm 0.013$ & $0.959 \pm 0.011$ & $0.968 \pm 0.011$ & $0.958 \pm 0.019$ & $0.963 \pm 0.010$\\
    $C_{20}$  &  $95.88 \pm 1.12$ & $0.951 \pm 0.019$ &  $0.962 \pm 0.013$ & $0.957 \pm 0.012$ & $0.966 \pm 0.011$ & $0.956 \pm 0.018$ & $0.961 \pm 0.011$\\\hline
    $D_4$  &  $95.45 \pm 1.38$ & $0.948 \pm 0.024$ & $0.957 \pm 0.017$ & $0.952 \pm 0.014$ &  $0.962 \pm 0.015$ & $0.952 \pm 0.023$ & $0.957 \pm 0.013$\\
    $D_8$  &  $96.37 \pm 0.95$ & $0.960 \pm 0.019$ & $0.964 \pm 0.014$ & $0.962 \pm 0.010$ & $0.968 \pm 0.012$ & $0.964 \pm 0.018$ & $0.966 \pm 0.009$\\
    $D_{16}$ & $\mathbf{96.56}\pm\mathbf{1.29}$ & $0.963 \pm 0.025$ & $0.965 \pm 0.014$ & $0.964 \pm 0.013$ & $0.969 \pm 0.012$ & $0.966 \pm 0.023$ & $0.967 \pm 0.012$\\
    $D_{20}$  &  $96.39 \pm 1.00$ & $0.959 \pm 0.018$ & $0.966 \pm 0.015$ & $0.962 \pm 0.010$ & $0.969 \pm 0.013$ & $0.962 \pm 0.017$ & $0.966 \pm 0.010$\\\hline
    \end{tabular}
\end{table*}

From Table~\ref{tab:frf_metrics}, it can be seen that the best test accuracy is achieved by the $D_{16}$ model, highlighted in bold. Indeed, while all equivariant models perform better than the standard CNN, the performance of the dihedral models is consistently better than for the cyclic models of equivalent order. 

For the cyclic models it can be observed that the largest change in performance comes from an increased FRI recall. For a binary classification problem, the recall of a class is defined as 
\begin{equation}
    {\rm Recall} = \frac{\rm TP}{\rm TP + FN}~,
\end{equation}
where TP indicates the number of true positives and FN indicates the number of false negatives. The recall therefore represents the fraction of all objects in that class which are correctly classified. Equivalently, the precision of the class is defined as
\begin{equation}
    {\rm Precision} = \frac{\rm TP}{\rm TP + FP}~.
\end{equation}
Consequently, if the recall of one class increases at the expense of the precision of the opposing class then it indicates that the opposing class is being disproportionately misclassified. However, in this case we can observe from Table~\ref{tab:frf_metrics} that the precision of the FRII class is also increasing, suggesting that the improvement in performance is due to a smaller number of FRI objects being misclassified as FRII. For the cyclic models there is a smaller but not equivalent improvement in FRII recall. This suggests that the cyclic model primarily reduces the misclassification of FRI objects as FRII, but does not equivalently reduce the misclassification of FRII as FRI. 

\begin{figure}
    \centering
    \includegraphics[width=0.5\textwidth]{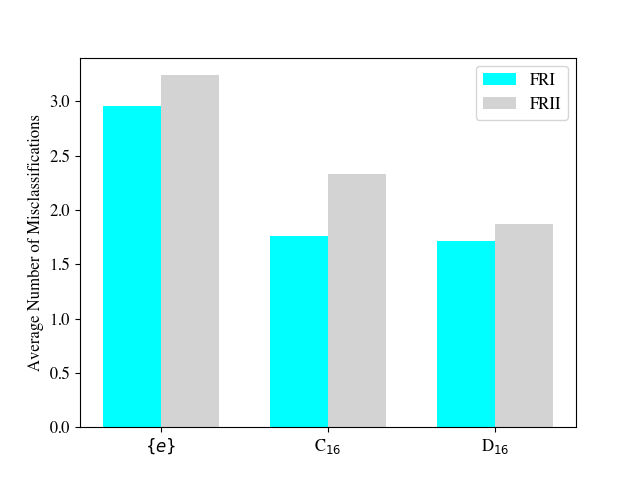}
    \caption{Average number of misclassifications for FRI (cyan) and FRII (grey) over all orientations and training repeats for the standard CNN, denoted $\{ e\}$, the $C_{16}$ CNN and the $D_{16}$ CNN, see Section~\ref{sec:performance} for details. \label{fig:misclass}}
\end{figure}

The dihedral models show a more even distribution of improvement across all metrics, indicating that there are more balanced reductions across both FRI and FRII misclassifications. This is illustrated in Figure~\ref{fig:misclass}, which shows the average number of misclassifications over all orientations and training repeats for the standard CNN, the $C_{16}$ CNN and the $D_{16}$ CNN for the reserved test set.

The test partition of the full \emph{Mirabest} data set contains 153 FRI and FRII-type sources labelled as both Confident and Uncertain, see Table~\ref{tab:data}. When using this combined test set the overall performance metrics of the networks considered in this work become accordingly lower due to the inclusion of the Uncertain sources. This is expected, not only because the Uncertain samples include edge cases that are more difficult
to classify but also because the assigned labels for these objects may not be fully accurate. However, the relative performance shows the same degree of improvement between the standard CNN, $\{e\}$, and the $D_{16}$ model, which have percentage accuracies of $82.59\pm1.41$ and $85.30\pm1.35$, respectively, when evaluated against this combined test set. 

We note that given the comparatively small size of the \emph{Mirabest$^{\ast}$} training set, these results may not generalise equivalently to other potentially larger data sets with different selection specifications and that additional validation should be performed when considering the use of group-equivariant convolutions for other classification problems.

\subsection{On the confidence of G-Steerable CNNs}
\label{sec:confidence}

Target class predictions for each test data sample are made by selecting the highest softmax probability, which provides a normalised version of the network output values. By using dropout as a Bayesian approximation, as demonstrated in \cite{mcdropout}, one is able to obtain a posterior distribution of network outputs for each test sample. This posterior distribution allows one to assess the degree of certainty with which a prediction is being made, i.e. if the distribution of outputs for a particular class is well-separated from those of other classes then the input is being classified with high confidence; however, if the distribution of outputs intersects those of other classes then, even though the softmax probability for a particular realisation may be high (even as high as unity), the overall distribution of softmax probabilities for that class may still fill the entire $[0,1]$ range, overlapping significantly with the distributions from other target classes. Such a circumstance denotes a low degree of model certainty in the softmax probability and therefore in the class prediction for that particular test sample. 

\begin{figure*}
\centerline{\includegraphics[width=0.5\textwidth]{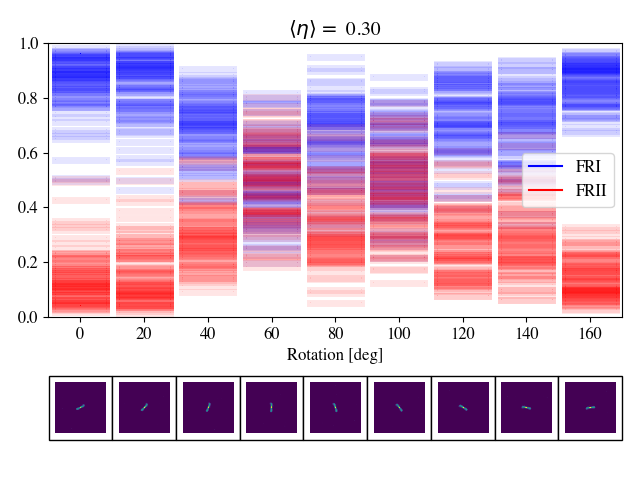}\qquad\includegraphics[width=0.5\textwidth]{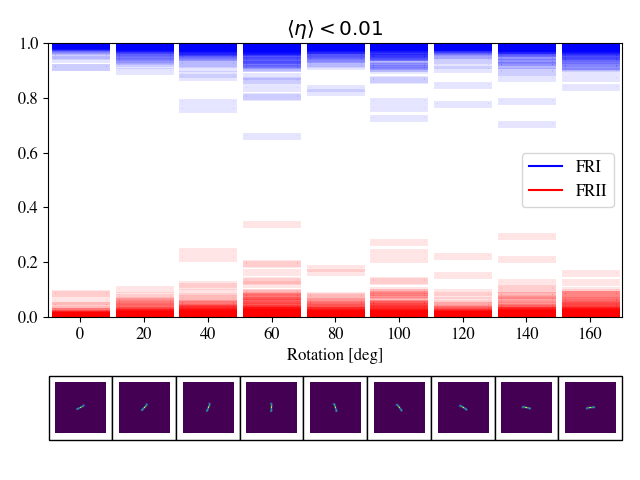}}
\caption{A scatter of 50 forward passes of the softmax output for the standard CNN (left)  and the $D_{16}$-equivariant CNN (right). The lower panel shows the rotated image of the test image. As indicated, the average overlap index for the standard CNN is $\langle \eta \rangle = 0.30$, and $\langle \eta \rangle < 0.01$ for the $D_{16}$-equivariant CNN. \label{fig:rotation}}
\end{figure*}

By re-enabling the dropout before the final fully-connected layer at test time, we estimate the predictive uncertainty of each model for the data samples in the reserved \emph{MiraBest$^{\ast}$} test set. With dropout enabled, we perform $T=50$ forward passes through the trained network for each sample in the test set. On each pass we recover $(x_t, y_t)$, where $x$ and $y$ are the softmax probabilities of FRI and FRII, respectively. An example of the results from this process can be seen in Figure~\ref{fig:rotation}, where we evaluate the trained model on a rotated version of the input image at discrete intervals of 20$^{\circ}$ in the range $[0^{\circ}, 180^{\circ})$ using a trained model for the standard CNN (left panel) and for the $D_{16}$-equivariant CNN (right panel). For each rotation angle, a distribution of softmax probabilities is obtained. In the case of the standard CNN it can be seen that, although the model classifies the source with high confidence when it is unrotated (0$^{\circ}$), the softmax probability distributions are not well-separated for the central image orientations, indicating that the model has a lower degree of confidence in the prediction being made in at these orientations. For the $D_{16}$-equivariant CNN it can be seen that in this particular test case the model has a high degree of confidence in its prediction for all orientations of the image. 

To represent the degree of uncertainty for each test sample quantitatively, we evaluate the degree of overlap in the distributions of softmax probabilities at a particular rotation angle using the distribution-free overlap index \citep{overlap}.  To do this, we calculate the local densities at position $z$ for each class using a Gaussian kernel density estimator, such that
\begin{eqnarray}
f_x(z) &=& \frac{1}{T} \sum_{t=1}^{T} { \frac{1}{\beta \sqrt{2\pi}} {\rm e}^{-(z - x_t)^2/2\beta^2} },\\
f_y(z) &=& \frac{1}{T} \sum_{t=1}^{T} { \frac{1}{\beta \sqrt{2\pi}} {\rm e}^{-(z - y_t)^2/2\beta^2} },
\end{eqnarray}
where $\beta=0.1$. We then use these local densities to calculate the overlap index, $\eta$, such that
\begin{equation}
    \eta = \sum_{i=1}^{N_z}{ {\rm min}\left[f_x(z_i), f_y(z_i)\right]\delta z },
\end{equation}
where $\left\{z_i\right\}_{i=1}^{N_z}$ covers the range zero to one in $N_z$ steps of size $\delta z$. For this work we assume $N_z = 100$. The resulting overlap index, $\eta$, varies between zero and one, with larger values indicating a higher degree of overlap and hence a lower degree of confidence.  

For each test sample we evaluate the overlap index over a range of rotations from 0$^{\circ}$ to 180$^{\circ}$ in increments of 20$^{\circ}$. We then calculate the average overlap index, $\langle \eta \rangle$, across these nine rotations. In Figure~\ref{fig:rotation} the value of this index can be seen above each plot: in this case, the standard CNN has $\langle \eta \rangle_{\{e\}} = 0.30$ and the $D_{16}$-equivariant CNN has $\langle \eta \rangle_{D_{16}} < 0.01$.

Of the 104 data samples in the reserved test set, $27.7\pm11.0$\% of objects show an improvement in average model confidence, i.e.  $\langle \eta \rangle_{\{e\}} - \langle \eta \rangle_{D_{16}} > 0.01$, when classified using the $D_{16}$-equivariant CNN compared to the standard CNN, $8.4\pm2.5$\% show a deterioration in average model confidence, i.e.  $\langle \eta \rangle_{D_{16}} - \langle \eta \rangle_{\{e\}} > 0.01$, and all other samples show no significant change in average model confidence, i.e.  $|\langle \eta \rangle_{\{e\}} - \langle \eta \rangle_{D_{16}}| < 0.01$. Mean values and uncertainties are determined from $\langle \eta \rangle$ values for all test samples evaluated using a pairwise comparison of 5 training realisations of the standard CNN and 5 training realisations of the $D_{16}$ CNN. 

Those objects that show an improvement in average model confidence are approximately evenly divided between FRI and  FRII type objects, whereas the objects that show a reduction in model confidence exhibit a weak preference for FRII. These results are discussed further in Section~\ref{sec:orientation}.

\section{Discussion}
\label{sec:discussion}

\subsection{Statistical distribution of radio galaxy orientations}
\label{sec:orientation}

Mathematically, $G$-steerable CNNs classify equivalence classes of images, as defined by the equivalence relation of a particular group, $G$, whereas conventional CNNs classify equivalence classes defined only by translations. Consequently, by using E(2)-equivalent convolutions the trained models assume that the statistics of extra-galactic astronomical images containing individual objects are expected to be invariant not only to translations but also to global rotations and reflections. Here we briefly review the literature in order to consider whether this assumption is robust and highlight the limitations that may result from it.

The orientation of radio galaxies, as defined by the direction of their jets, is thought to be determined by the angular momentum axis of the super-massive black hole within the host galaxy. A number of studies have looked for evidence of preferred jet alignment directions in populations of radio galaxies, as this has been proposed to be a potential consequence of angular momentum transfer during galaxy formation \citep[e.g.][]{white1984, codis2018, kraljic2020}, or alternatively it could be caused by large-scale filamentary structures in the cosmic web giving rise to preferential merger directions \citep[see e.g.][]{kartaltepe2008} that might result in jet alignment for radio galaxies formed during mergers \citep[e.g.][]{croton2006, chiaberge2015}. The observational evidence for both remains a subject of discussion in the literature.

\cite{taylorpresh2016} found a local alignment  of radio galaxies in the ELAIS~N1 field on scales $<1^{\circ}$ using observations from the Giant Metrewave Radio Telescope (GMRT) at 610\,MHz. Local alignments were also  reported by \cite{contigiani2017} who reported evidence ($>2\sigma$) of local alignment on scales of $\sim 2.5^{\circ}$ among radio sources from the FIRST survey using a much larger sample of radio galaxies, catalogued by the radio galaxy zoo project. A similar local alignment was also reported by \cite{panwar2020} using data from the FIRST survey. Using a sample of 7555 double-lobed radio galaxies from the LOFAR Sky Survey \citep[LoTSS;][]{lotss} at 150\,MHz, \cite{osinga2020} concluded that a statistical deviation from purely random distributions of orientation as a function of projected distance was caused by systematics introduced by the brightest objects and did not persist when redshift information was taken into account. However, the study also suggested that larger samples of radio galaxies should be used to confirm this result.

Whilst these results may suggest tentative evidence for spatial correlations of radio galaxy orientations in local large-scale structure, they do not provide any information on whether these orientations differ between classes of radio galaxy, i.e. the equivalence classes considered here. Moreover, the large spatial distribution and comparatively small number of galaxies that form the training set used in this work mean that even spatial correlation effects would be unlikely to be significant for the data set used here. However, the results of \cite{taylorpresh2016, contigiani2017, panwar2020} suggest that care should be taken in this assumption if data sets are compiled from only small spatial regions. 

In Section~\ref{sec:convergence}, we found that the largest improvement in performance was seen when using dihedral, $D_N$, models. We suggest that this improvement over cyclic, $C_N$, models is due to image reflections accounting for chirality, in addition to orientations on the celestial sphere which are represented by the cyclic group. Galactic chirality has previously been considered for populations of star-forming, or normal, galaxies \citep[see e.g.][]{slosar2009, shamir2020}, as the spiral structure of star-forming galaxies means that such objects can be considered to be enantiomers, i.e. their mirror images are not superimposable \citep{chirality}. It has been suggested that a small asymmetry exists in the number of clockwise versus anti-clockwise star-forming galaxy spins \citep{shamir2020}. As far as the authors are aware there have been no similar studies considering the chirality of radio galaxies. However, a simple example of such chirality for radio galaxies might include the case where relativistic boosting causes one jet of a radio galaxy to appear brighter than the other due to an inclination relative to the line of sight. Since the dominance of a particular orientation relative to the line of sight should be unbiased then this would imply a global equivariance to reflection. Since the dihedral ($D_N$) models used in this work are insensitive to chirality, the results in Section~\ref{sec:convergence} suggest that the radio galaxies in the training sample used here do not have a significant degree of preferred chirality. Whilst this does not itself validate the assumption of global reflection invariance, in the absence of evidence to the contrary from the literature we suggest that it is unlikely to be significant for the data sample used in this work.

From the perspective of classification, equivariance to reflections implies that inference should be independent of reflections of the input. For FR~I and FR~II radio galaxy classification, incorporating such information into a classification scheme may be important more generally: the unified picture of radio galaxies holds that both FR~I and FR~II, as well as many other classifications of active galactic nuclei (AGN) such as quasars, QSOs (quasi-stellar objects), blazars, BL Lac objects, Seyfert galaxies etc., are in fact defined by orientation-dependent observational differences, rather than intrinsic physical distinctions \citep{urry2004}. 

Consequently, under the assumptions of global rotational and reflection invariance, the possibility of a classification model providing different output classifications for the same test sample at different orientations is problematic. Furthermore, the degree of model confidence in a classification should also not vary significantly as a function of sample orientation, i.e. if a galaxy is confidently classified at one particular orientation then it should be approximately equally confidently classified at all other orientations. If this is not the case, as shown for the standard CNN in Figure~\ref{fig:rotation} (left), then it indicates a preferred orientation in the model weights for a given outcome, inconsistent with the expected statistics of the true source population. Such inconsistencies might be expected to result in biased samples being extracted from survey data. 

In this context it is then not only the average degree of model confidence that is important as a function of sample rotation, as quantified by the value of $\langle \eta \rangle$ in Section~\ref{sec:confidence}, but also the stability of the $\eta$ index as a function of rotation, i.e. a particular test sample should be classified at a consistent degree of confidence as a function of orientation, whether that confidence is low or high. To evaluate the stability of the predictive confidence as a function of orientation, we examine the variance of the $\eta$ index as a function of rotation. For the \emph{MiraBest$^{\ast}$} reserved test set we find that approximately 30\% of the test samples show a reduction of more than 0.01 in the standard deviation of their overlap index as a function of rotation, with 17\% showing a reduction of more than 0.05. Conversely approximately 8\% of test samples show an increase of $> 0.01$ and 4\% samples show an increase of $> 0.05$. In a similar manner to the results for average model confidence given in Section~\ref{sec:confidence}, those objects that show a reduction in their variance, i.e. an improvement in the consistency of prediction as a function of rotation, are evenly balanced between the two classes; however, those objects showing a strong improvement of $>0.05$ are preferentially FRI type objects.


\subsection{Comment on Capsule Networks}
\label{sec:capsule}

The use of capsule networks \citep{sabour2017} for radio galaxy classification was investigated by \cite{lukic2018}. Capsule networks aim to separate the orientation (typically referred to as the viewpoint or pose in the context of capsule networks) of an object from its nature, i.e. class, by encoding the output of their layers as tuples incorporating both a \emph{pose vector} and an activation. The purpose of this approach is to focus on the linear hierarchical relationships in the data and remove sensitivity to orientation; however, as described by \cite{lenssen2018}, general capsule networks do not guarantee particular group equivariances and therefore cannot completely disentangle orientation from feature data. It is perhaps partly for this reason that \cite{lukic2018} found that capsule networks offered no significant advantage over standard CNNs for the radio galaxy classification problem addressed in that work. 

In Section~\ref{sec:results}, we found that not only is the test performance improved by the use of equivariant CNNs, but that equivariant networks also converge more rapidly. For image data, a standard CNN enables generalization over classes of translated images, which provides an advantage over the use of an MLP, where every image must be considered individually. $G$-steerable CNNs extend this behaviour to include additional equivalences, further improving generalization. This additional equivariance enhances the data efficiency of the learning algorithm because it means that every image is no longer an individual data point but instead a representative of its wider equivalence group. Consequently, unlike capsule networks, the equivalence groups being classified by a $G$-steerable CNN are specified a priori, rather than the orientations of individual samples being learned during training. Whilst this creates additional capacity in the network for learning intra-class differences that are insensitive to the specified equivalences, it does not provide  the information on orientation of individual samples that is provided as an output by capsule networks. 

\cite{lenssen2018} combined group-equivariant convolutions with capsule networks in order to output information on both classification and pose, although they note that a limitation of this combined approach is that arbitrary pose information is no longer available, but is instead  limited to the elements of the equivariant group. For radio astronomy, where radio galaxy orientations are expected to be extracted from images at a precision that is limited by the observational constraints of the data, it is unlikely that pose information limited to the elements of a low-order finite group, $G < E(2)$, is sufficient for further analysis. However, given particular sets of observational and physical constraints or specifications it is possible that such an approach may become useful at some limiting order. Alternatively, pose information might be used to specify a prior for a secondary processing step that refines a measurement of orientation.

\subsection{Local vs Global Equivariance}
\label{sec:local}

By design, the final features used for classification in equivariant CNNs do not include any information about the global orientation or chirality of an input image; however, this can also mean that they are insensitive to local equivariances in the image, when these might in fact be useful for classification. The hierarchical nature of convolutional networks can be used to mitigate against this, as kernels corresponding to earlier layers in a network will have a smaller, more local, footprint on the input image and therefore be sensitive to a different scale of feature than those from deeper layers which encompass larger-scale information. Therefore, by changing the degree of equivariance as a function of layer depth one can control the degree to which local equivariance is enforced. \cite{weilercesa2019} refer to this practice as \emph{group restriction} and find that it is beneficial when classifying data sets that possess symmetries on a local scale but not on a global scale, such as the CIFAR and unrotated MNIST datasets. Conversely, the opposite situation may also be true, where no symmetry is present on a local scale, but the data are statistically invariant on a global scale. In this case the reverse may be done and, rather than restricting the representation of the feature data to reduce the degree of equivariance, one might expand the domain of the representation at a particular layer depth in order to reflect a global equivariance. 
\begin{figure}
\centerline{\includegraphics[width=0.5\textwidth]{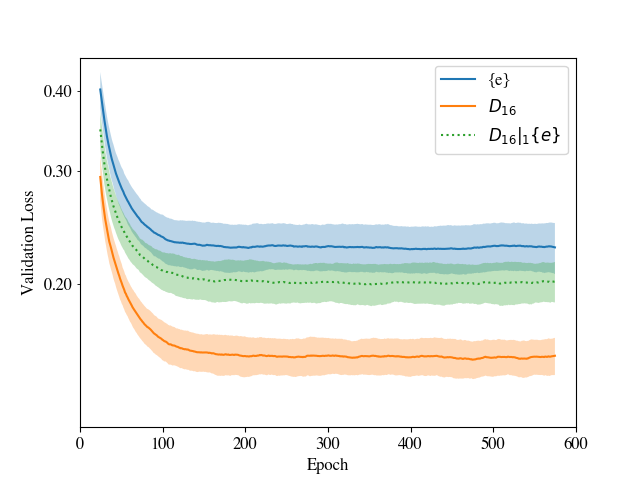}}
\caption{Validation losses during the training of the standard CNN, denoted $\{e\}$ (blue), the $D_{16}$ CNN (orange), and the restricted $D_{N}|_1\{e\}$ CNN (green; dashed) for the \emph{MiraBest$^{\ast}$} data set. Plots show mean and standard deviation over five training repeats. \label{fig:restricted}}
\end{figure}

We investigate the effect of group restriction by using a $D_{N}|_1\{e\}$ restricted version of the LeNet architecture, i.e. the first layer is $D_{N}$ equivariant and the second convolutional layer is a standard convolution. Using $N=16$, the loss curve for this restricted architecture relative to the unrestricted $D_{16}$ equivariant CNN is shown in Figure~\ref{fig:restricted}. From the figure it can be seen that while exploiting local symmetries gives an improved performance over the standard CNN, the performance of the group restricted model is significantly poorer than that of the full $D_{16}$ CNN. This result suggests that although local symmetries are present in the data, it is the global symmetries of the population that result in the larger performance gain for the radio galaxy data set.

\subsection{Note on hyper-parameter tuning}
\label{sec:hp}

In Section~\ref{sec:results} we found that the $N=16$ cyclic and dihedral models were preferred over the higher order $N=20$ models. This may seem counter-intuitive as one might assume that for truly rotationally invariant data sets the performance would converge to a limiting value as the order increased, rather than finding a minimum at some discrete point. Consequently, we note that the observed minimum at $N=16$ might not represent a true property of the data set but instead represent a limitation caused by discretisation artifacts from rotation of convolution kernels with small support, in this case $k=5$, see Table~\ref{tab:lenet} \citep{weilercesa2019}. These same discretisation errors may also account in part for the small oscillation in validation error as a function of group order seen in Figure~\ref{fig:err_order}. Consequently, while no additional hyper-parameter tuning has been performed for any of the networks used in this work, we note that kernel size is potentially one hyper-parameter that could be tuned as a function of group order, $N$, and that such tuning might lead to further improvements in performance for higher orders.

\section{Conclusions}
\label{sec:conc}

In this work, we have demonstrated that the use of even low-order group-equivariant convolutions results in a performance improvement over standard convolutions for the radio galaxy classification problem considered here, without additional hyper-parameter tuning. We have shown that both cyclic and dihedral equivariant models converge to lower validation loss values during training and provide improved validation errors. We attribute this improvement to the increased capacity of the equivariant networks for learning hierarchical features specific to classification, when additional capacity for encoding redundant feature information at multiple orientations is no longer required, hence reducing intra-class variability. 

We have shown that for the simple network architecture and training set considered here, a $D_{16}$ equivariant model results in the best test performance using a reserved test set. We suggest that the improvement of the dihedral over the cyclic models is due to an insensitivity to - and therefore lack of preferred - chirality in the data, and that further improvements in performance might be gained from tuning the size of the kernels in the convolutional layers according to the order of the equivalence group. We find that cyclic models predominantly reduce the misclassification of FRI type radio galaxies, whereas dihedral models reduce misclassifications for both FRI and FRII type galaxies. 

By using the MC Dropout Bayesian approximation method, we have shown that the improved performance observed for the $D_{16}$ model compared to the standard CNN is reflected in the model confidence as a function of rotation. Using the reserved test set, we have quantified this difference in confidence using the overlap between predictive probability distributions of different target classes, as encapsulated in the distribution free overlap index parameter, $\eta$. We find that not only is average model confidence improved when using the equivariant model, but also that the consistency of model confidence as a function of image orientation is improved. We emphasise the importance of such consistency for applications of CNN-based classification in order to avoid biases in samples being extracted from future survey data.

Whilst the results presented here are encouraging, we note that this work addresses a specific classification problem in radio astronomy and the  method used here may not result in equivalent improvements when applied to other areas of astronomical image classification using different data sets or network architectures. In particular, the assumptions of global rotational and reflectional invariance are strong assumptions, which may not apply to all data sets. As described in Section~\ref{sec:orientation}, data sets extracted from localised regions of the sky may be particularly vulnerable to biases when using this method and the properties of the \emph{MiraBest$^{\ast}$} data set used in this work may not generalise to all other data sets or classification problems. We note that this is true for all CNNs benchmarked against finite data sets and users should be aware that additional validation should be performed before models are deployed on new test data, as biases arising from data selection may be reflected in biases in classifier performance \cite[see e.g.][]{wu2018, tang2019frdeep, walmsley2020}. However, in conclusion, we echo the expectation of \cite{weilercesa2019}, that equivariant CNNs may soon become a common choice for morphological classification in fields like astronomy, where symmetries may be present in the data, and note that the overhead in constructing such networks is now minimal due to the emergence of standardised libraries such as {\tt e2cnn}. Future work will need to address the optimal architectures and hyper-parameter choices for such models as specific applications evolve.

\section*{Acknowledgements}

This work makes extensive use of the {\tt e2cnn} library: \url{github.com/QUVA-Lab/e2cnn}.
AMS gratefully acknowledges support from an Alan Turing Institute AI Fellowship EP/V030302/1. FP gratefully acknowledges support from STFC and IBM through the iCASE studentship ST/P006795/1.

\section*{Data Availability}

Code for this work is publicly available on Github at the following  address: \url{github.com/as595/E2CNNRadGal}. The MiraBest data set is available on Zenodo: 10.5281/zenodo.4288837.



\bibliographystyle{mnras}
\bibliography{e2_paper} 

\begin{thebibliography}{}
\makeatletter
\relax
\def\mn@urlcharsother{\let\do\@makeother \do\$\do\&\do\#\do\^\do\_\do\%\do\~}
\def\mn@doi{\begingroup\mn@urlcharsother \@ifnextchar [ {\mn@doi@}
  {\mn@doi@[]}}
\def\mn@doi@[#1]#2{\def\@tempa{#1}\ifx\@tempa\@empty \href
  {http://dx.doi.org/#2} {doi:#2}\else \href {http://dx.doi.org/#2} {#1}\fi
  \endgroup}
\def\mn@eprint#1#2{\mn@eprint@#1:#2::\@nil}
\def\mn@eprint@arXiv#1{\href {http://arxiv.org/abs/#1} {{\tt arXiv:#1}}}
\def\mn@eprint@dblp#1{\href {http://dblp.uni-trier.de/rec/bibtex/#1.xml}
  {dblp:#1}}
\def\mn@eprint@#1:#2:#3:#4\@nil{\def\@tempa {#1}\def\@tempb {#2}\def\@tempc
  {#3}\ifx \@tempc \@empty \let \@tempc \@tempb \let \@tempb \@tempa \fi \ifx
  \@tempb \@empty \def\@tempb {arXiv}\fi \@ifundefined
  {mn@eprint@\@tempb}{\@tempb:\@tempc}{\expandafter \expandafter \csname
  mn@eprint@\@tempb\endcsname \expandafter{\@tempc}}}

\bibitem[\protect\citeauthoryear{{Abazajian} et~al.,}{{Abazajian}
  et~al.}{2009}]{sdssdr7}
{Abazajian} K.~N.,  et~al., 2009, \mn@doi [\apjs]
  {10.1088/0067-0049/182/2/543}, \href
  {https://ui.adsabs.harvard.edu/abs/2009ApJS..182..543A} {182, 543}

\bibitem[\protect\citeauthoryear{Alger et~al.,}{Alger et~al.}{2018}]{alger2018}
Alger M.~J.,  et~al., 2018, \mn@doi [Monthly Notices of the Royal Astronomical
  Society] {10.1093/mnras/sty1308}, 478, 5547

\bibitem[\protect\citeauthoryear{Aniyan \& Thorat}{Aniyan \&
  Thorat}{2017}]{aniyan2017}
Aniyan A.~K.,  Thorat K.,  2017, \mn@doi [The Astrophysical Journal Supplement
  Series] {10.3847/1538-4365/aa7333}

\bibitem[\protect\citeauthoryear{{Banfield} et~al.,}{{Banfield}
  et~al.}{2015}]{rgz}
{Banfield} J.~K.,  et~al., 2015, \mn@doi [\mnras] {10.1093/mnras/stv1688},
  \href {https://ui.adsabs.harvard.edu/abs/2015MNRAS.453.2326B} {453, 2326}

\bibitem[\protect\citeauthoryear{Beardsley et~al.,}{Beardsley
  et~al.}{2019}]{Beardsley2019}
Beardsley A.~P.,  et~al., 2019, \mn@doi [Publications of the Astronomical
  Society of Australia] {10.1017/pasa.2019.41}

\bibitem[\protect\citeauthoryear{{Becker}, {White}  \& {Helfand}}{{Becker}
  et~al.}{1995}]{FIRST}
{Becker} R.~H.,  {White} R.~L.,   {Helfand} D.~J.,  1995, \mn@doi [\apj]
  {10.1086/176166}, \href
  {https://ui.adsabs.harvard.edu/abs/1995ApJ...450..559B} {450, 559}

\bibitem[\protect\citeauthoryear{{Best} \& {Heckman}}{{Best} \&
  {Heckman}}{2012}]{bestheckman}
{Best} P.~N.,  {Heckman} T.~M.,  2012, \mn@doi [\mnras]
  {10.1111/j.1365-2966.2012.20414.x}, \href
  {https://ui.adsabs.harvard.edu/abs/2012MNRAS.421.1569B} {421, 1569}

\bibitem[\protect\citeauthoryear{{Bowles}, {Scaife}, {Porter}, {Tang}  \&
  {Bastien}}{{Bowles} et~al.}{2020}]{bowles2020}
{Bowles} M.,  {Scaife} A. M.~M.,  {Porter} F.,  {Tang} H.,   {Bastien} D.~J.,
  2020, arXiv e-prints, \href
  {https://ui.adsabs.harvard.edu/abs/2020arXiv201201248B} {p. arXiv:2012.01248}

\bibitem[\protect\citeauthoryear{{Capozziello} \& {Lattanzi}}{{Capozziello} \&
  {Lattanzi}}{2005}]{chirality}
{Capozziello} S.,  {Lattanzi} A.,  2005, arXiv e-prints, \href
  {https://ui.adsabs.harvard.edu/abs/2005physics...9144C} {p. physics/0509144}

\bibitem[\protect\citeauthoryear{{Chiaberge}, {Gilli}, {Lotz}  \&
  {Norman}}{{Chiaberge} et~al.}{2015}]{chiaberge2015}
{Chiaberge} M.,  {Gilli} R.,  {Lotz} J.~M.,   {Norman} C.,  2015, \mn@doi
  [\apj] {10.1088/0004-637X/806/2/147}, \href
  {https://ui.adsabs.harvard.edu/abs/2015ApJ...806..147C} {806, 147}

\bibitem[\protect\citeauthoryear{{Codis}, {Jindal}, {Chisari}, {Vibert},
  {Dubois}, {Pichon}  \& {Devriendt}}{{Codis} et~al.}{2018}]{codis2018}
{Codis} S.,  {Jindal} A.,  {Chisari} N.~E.,  {Vibert} D.,  {Dubois} Y.,
  {Pichon} C.,   {Devriendt} J.,  2018, \mn@doi [\mnras]
  {10.1093/mnras/sty2567}, \href
  {https://ui.adsabs.harvard.edu/abs/2018MNRAS.481.4753C} {481, 4753}

\bibitem[\protect\citeauthoryear{{Cohen} \& {Welling}}{{Cohen} \&
  {Welling}}{2016}]{cohenwelling2016}
{Cohen} T.~S.,  {Welling} M.,  2016, arXiv e-prints, \href
  {https://ui.adsabs.harvard.edu/abs/2016arXiv160207576C} {p. arXiv:1602.07576}

\bibitem[\protect\citeauthoryear{{Condon}, {Cotton}, {Greisen}, {Yin},
  {Perley}, {Taylor}  \& {Broderick}}{{Condon} et~al.}{1998}]{NVSS}
{Condon} J.~J.,  {Cotton} W.~D.,  {Greisen} E.~W.,  {Yin} Q.~F.,  {Perley}
  R.~A.,  {Taylor} G.~B.,   {Broderick} J.~J.,  1998, \mn@doi [\aj]
  {10.1086/300337}, \href
  {https://ui.adsabs.harvard.edu/abs/1998AJ....115.1693C} {115, 1693}

\bibitem[\protect\citeauthoryear{{Contigiani} et~al.,}{{Contigiani}
  et~al.}{2017}]{contigiani2017}
{Contigiani} O.,  et~al., 2017, \mn@doi [\mnras] {10.1093/mnras/stx1977}, \href
  {https://ui.adsabs.harvard.edu/abs/2017MNRAS.472..636C} {472, 636}

\bibitem[\protect\citeauthoryear{{Croton} et~al.,}{{Croton}
  et~al.}{2006}]{croton2006}
{Croton} D.~J.,  et~al., 2006, \mn@doi [\mnras]
  {10.1111/j.1365-2966.2005.09675.x}, \href
  {https://ui.adsabs.harvard.edu/abs/2006MNRAS.365...11C} {365, 11}

\bibitem[\protect\citeauthoryear{Dieleman, Willett  \& Dambre}{Dieleman
  et~al.}{2015}]{dieleman2015}
Dieleman S.,  Willett K.~W.,   Dambre J.,  2015, \mn@doi [Monthly Notices of
  the Royal Astronomical Society] {10.1093/mnras/stv632}, 450, 1441

\bibitem[\protect\citeauthoryear{{Dieleman}, {De Fauw}  \&
  {Kavukcuoglu}}{{Dieleman} et~al.}{2016}]{dieleman2016}
{Dieleman} S.,  {De Fauw} J.,   {Kavukcuoglu} K.,  2016, arXiv e-prints, \href
  {https://ui.adsabs.harvard.edu/abs/2016arXiv160202660D} {p. arXiv:1602.02660}

\bibitem[\protect\citeauthoryear{{Fanaroff} \& {Riley}}{{Fanaroff} \&
  {Riley}}{1974}]{fr1974}
{Fanaroff} B.~L.,  {Riley} J.~M.,  1974, \mn@doi [\mnras]
  {10.1093/mnras/167.1.31P}, \href
  {https://ui.adsabs.harvard.edu/abs/1974MNRAS.167P..31F} {167, 31P}

\bibitem[\protect\citeauthoryear{{Gal} \& {Ghahramani}}{{Gal} \&
  {Ghahramani}}{2015}]{mcdropout}
{Gal} Y.,  {Ghahramani} Z.,  2015, arXiv e-prints, \href
  {https://ui.adsabs.harvard.edu/abs/2015arXiv150602142G} {p. arXiv:1506.02142}

\bibitem[\protect\citeauthoryear{{Gens} \& {Domingos}}{{Gens} \&
  {Domingos}}{2014}]{gensdomingos}
{Gens} R.,  {Domingos} P.~M.,  2014, in {Advances in Neural Information
  Processing Systems}. Curran Associates, Inc., pp 2537--2545

\bibitem[\protect\citeauthoryear{{Gheller}, {Vazza}  \& {Bonafede}}{{Gheller}
  et~al.}{2018}]{gheller2018}
{Gheller} C.,  {Vazza} F.,   {Bonafede} A.,  2018, \mn@doi [\mnras]
  {10.1093/mnras/sty2102}, \href
  {https://ui.adsabs.harvard.edu/abs/2018MNRAS.480.3749G} {480, 3749}

\bibitem[\protect\citeauthoryear{{Ginsburg} et~al.,}{{Ginsburg}
  et~al.}{2019}]{astroquery}
{Ginsburg} A.,  et~al., 2019, \mn@doi [\aj] {10.3847/1538-3881/aafc33}, \href
  {https://ui.adsabs.harvard.edu/abs/2019AJ....157...98G} {157, 98}

\bibitem[\protect\citeauthoryear{Jarvis et~al.,}{Jarvis
  et~al.}{2016}]{Jarvis2016}
Jarvis M.~J.,  et~al., 2016, in Proceedings of Science. ,
  \mn@doi{10.22323/1.277.0006}

\bibitem[\protect\citeauthoryear{Johnston et~al.,}{Johnston
  et~al.}{2008}]{Johnston2008}
Johnston S.,  et~al., 2008, \mn@doi [Experimental Astronomy]
  {10.1007/s10686-008-9124-7}

\bibitem[\protect\citeauthoryear{Kartaltepe, Ebeling, Ma  \&
  Donovan}{Kartaltepe et~al.}{2008}]{kartaltepe2008}
Kartaltepe J.~S.,  Ebeling H.,  Ma C.~J.,   Donovan D.,  2008, \mn@doi [Monthly
  Notices of the Royal Astronomical Society]
  {10.1111/j.1365-2966.2008.13620.x}, 389, 1240

\bibitem[\protect\citeauthoryear{{Kingma} \& {Ba}}{{Kingma} \&
  {Ba}}{2014}]{adam}
{Kingma} D.~P.,  {Ba} J.,  2014, arXiv e-prints, \href
  {https://ui.adsabs.harvard.edu/abs/2014arXiv1412.6980K} {p. arXiv:1412.6980}

\bibitem[\protect\citeauthoryear{{Kraljic}, {Dav{\'e}}  \& {Pichon}}{{Kraljic}
  et~al.}{2020}]{kraljic2020}
{Kraljic} K.,  {Dav{\'e}} R.,   {Pichon} C.,  2020, \mn@doi [\mnras]
  {10.1093/mnras/staa250}, \href
  {https://ui.adsabs.harvard.edu/abs/2020MNRAS.493..362K} {493, 362}

\bibitem[\protect\citeauthoryear{Krizhevsky, Sutskever  \& Hinton}{Krizhevsky
  et~al.}{2012}]{lenetdropout}
Krizhevsky A.,  Sutskever I.,   Hinton G.~E.,  2012, in Pereira F.,  Burges C.
  J.~C.,  Bottou L.,   Weinberger K.~Q.,  eds, , Advances in Neural Information
  Processing Systems 25.
Curran Associates, Inc., pp 1097--1105

\bibitem[\protect\citeauthoryear{LeCun, Bottou, Bengio  \& Haffner}{LeCun
  et~al.}{1998}]{lenet}
LeCun Y.,  Bottou L.,  Bengio Y.,   Haffner P.,  1998, Proceedings of the IEEE,
  86, 2278

\bibitem[\protect\citeauthoryear{{LeCun}, {Bottou}, {Orr}  \&
  {M\"{u}ller}}{{LeCun} et~al.}{2012}]{tricks}
{LeCun} Y.,  {Bottou} L.,  {Orr} G.,   {M\"{u}ller} K.,  2012, in {Neural
  Networks: Tricks of the Trade. Lecture Notes in Computer Science}. ,
  \mn@doi{10.1007/978-3-642-35289-8_3}

\bibitem[\protect\citeauthoryear{{Lenc} \& {Vedaldi}}{{Lenc} \&
  {Vedaldi}}{2014}]{lencvedaldi}
{Lenc} K.,  {Vedaldi} A.,  2014, arXiv e-prints, \href
  {https://ui.adsabs.harvard.edu/abs/2014arXiv1411.5908L} {p. arXiv:1411.5908}

\bibitem[\protect\citeauthoryear{{Lenssen}, {Fey}  \& {Libuschewski}}{{Lenssen}
  et~al.}{2018}]{lenssen2018}
{Lenssen} J.~E.,  {Fey} M.,   {Libuschewski} P.,  2018, arXiv e-prints, \href
  {https://ui.adsabs.harvard.edu/abs/2018arXiv180605086L} {p. arXiv:1806.05086}

\bibitem[\protect\citeauthoryear{{Lukic}, {Br{\"u}ggen}, {Banfield}, {Wong},
  {Rudnick}, {Norris}  \& {Simmons}}{{Lukic} et~al.}{2018}]{lukic2018}
{Lukic} V.,  {Br{\"u}ggen} M.,  {Banfield} J.~K.,  {Wong} O.~I.,  {Rudnick} L.,
   {Norris} R.~P.,   {Simmons} B.,  2018, \mn@doi [\mnras]
  {10.1093/mnras/sty163}, \href
  {https://ui.adsabs.harvard.edu/abs/2018MNRAS.476..246L} {476, 246}

\bibitem[\protect\citeauthoryear{{McGlynn}, {Scollick}  \& {White}}{{McGlynn}
  et~al.}{1998}]{skyview}
{McGlynn} T.,  {Scollick} K.,   {White} N.,  1998, in {McLean} B.~J.,
  {Golombek} D.~A.,  {Hayes} J. J.~E.,   {Payne} H.~E.,  eds, New Horizons from
  Multi-Wavelength Sky Surveys.

\bibitem[\protect\citeauthoryear{Miraghaei \& Best}{Miraghaei \&
  Best}{2017}]{mirabest2017}
Miraghaei H.,  Best P.~N.,  2017, \mn@doi [Monthly Notices of the Royal
  Astronomical Society] {10.1093/mnras/stx007}, 466, 4346

\bibitem[\protect\citeauthoryear{{Osinga} et~al.,}{{Osinga}
  et~al.}{2020}]{osinga2020}
{Osinga} E.,  et~al., 2020, \mn@doi [\aap] {10.1051/0004-6361/202037680}, \href
  {https://ui.adsabs.harvard.edu/abs/2020A&A...642A..70O} {642, A70}

\bibitem[\protect\citeauthoryear{{Panwar}, {Prabhakar}, {Sandhu}, {Wadadekar}
  \& {Jain}}{{Panwar} et~al.}{2020}]{panwar2020}
{Panwar} M.,  {Prabhakar} {Sandhu} P.~K.,  {Wadadekar} Y.,   {Jain} P.,  2020,
  \mn@doi [\mnras] {10.1093/mnras/staa2975}, \href
  {https://ui.adsabs.harvard.edu/abs/2020MNRAS.499.1226P} {499, 1226}

\bibitem[\protect\citeauthoryear{Pastore \& Calcagnì}{Pastore \&
  Calcagnì}{2019}]{overlap}
Pastore M.,  Calcagnì A.,  2019, \mn@doi [Frontiers in Psychology]
  {10.3389/fpsyg.2019.01089}, 10, 1089

\bibitem[\protect\citeauthoryear{Ren, He, Girshick  \& Sun}{Ren
  et~al.}{2015}]{fasterrcnn}
Ren S.,  He K.,  Girshick R.,   Sun J.,  2015, in Advances in neural
  information processing systems. pp 91--99

\bibitem[\protect\citeauthoryear{{Sabour}, {Frosst}  \& {E Hinton}}{{Sabour}
  et~al.}{2017}]{sabour2017}
{Sabour} S.,  {Frosst} N.,   {E Hinton} G.,  2017, arXiv e-prints, \href
  {https://ui.adsabs.harvard.edu/abs/2017arXiv171009829S} {p. arXiv:1710.09829}

\bibitem[\protect\citeauthoryear{{Shamir}}{{Shamir}}{2020}]{shamir2020}
{Shamir} L.,  2020, \mn@doi [\pasa] {10.1017/pasa.2020.46}, \href
  {https://ui.adsabs.harvard.edu/abs/2020PASA...37...53S} {37, e053}

\bibitem[\protect\citeauthoryear{{Shimwell} et~al.,}{{Shimwell}
  et~al.}{2019}]{lotss}
{Shimwell} T.~W.,  et~al., 2019, \mn@doi [\aap] {10.1051/0004-6361/201833559},
  \href {https://ui.adsabs.harvard.edu/abs/2019A&A...622A...1S} {622, A1}

\bibitem[\protect\citeauthoryear{{Slosar} et~al.,}{{Slosar}
  et~al.}{2009}]{slosar2009}
{Slosar} A.,  et~al., 2009, \mn@doi [\mnras]
  {10.1111/j.1365-2966.2008.14127.x}, \href
  {https://ui.adsabs.harvard.edu/abs/2009MNRAS.392.1225S} {392, 1225}

\bibitem[\protect\citeauthoryear{Tang}{Tang}{2019}]{tang2019frdeep}
Tang H.,  2019, FR-DEEP, \url{https://github.com/HongmingTang060313/FR-DEEP}

\bibitem[\protect\citeauthoryear{{Tang}, {Scaife}  \& {Leahy}}{{Tang}
  et~al.}{2019}]{hmtnet}
{Tang} H.,  {Scaife} A.~M.~M.,   {Leahy} J.~P.,  2019, \mn@doi [\mnras]
  {10.1093/mnras/stz1883}, \href
  {https://ui.adsabs.harvard.edu/abs/2019MNRAS.488.3358T} {488, 3358}

\bibitem[\protect\citeauthoryear{{Taylor} \& {Jagannathan}}{{Taylor} \&
  {Jagannathan}}{2016}]{taylorpresh2016}
{Taylor} A.~R.,  {Jagannathan} P.,  2016, \mn@doi [\mnras]
  {10.1093/mnrasl/slw038}, \href
  {https://ui.adsabs.harvard.edu/abs/2016MNRAS.459L..36T} {459, L36}

\bibitem[\protect\citeauthoryear{{Urry}}{{Urry}}{2004}]{urry2004}
{Urry} C.,  2004, in {Richards} G.~T.,  {Hall} P.~B.,  eds,  Astronomical
  Society of the Pacific Conference Series Vol. 311, AGN Physics with the Sloan
  Digital Sky Survey. p.~49 (\mn@eprint {arXiv} {astro-ph/0312545})

\bibitem[\protect\citeauthoryear{Van~Haarlem et~al.,}{Van~Haarlem
  et~al.}{2013}]{VanHaarlem2013}
Van~Haarlem M.~P.,  et~al., 2013, {LOFAR: The low-frequency array},
  \mn@doi{10.1051/0004-6361/201220873}

\bibitem[\protect\citeauthoryear{{Walmsley} et~al.,}{{Walmsley}
  et~al.}{2020}]{walmsley2020}
{Walmsley} M.,  et~al., 2020, \mn@doi [\mnras] {10.1093/mnras/stz2816}, \href
  {https://ui.adsabs.harvard.edu/abs/2020MNRAS.491.1554W} {491, 1554}

\bibitem[\protect\citeauthoryear{{Weiler} \& {Cesa}}{{Weiler} \&
  {Cesa}}{2019}]{weilercesa2019}
{Weiler} M.,  {Cesa} G.,  2019, arXiv e-prints, \href
  {https://ui.adsabs.harvard.edu/abs/2019arXiv191108251W} {p. arXiv:1911.08251}

\bibitem[\protect\citeauthoryear{{Weiler}, {Hamprecht}  \& {Storath}}{{Weiler}
  et~al.}{2017}]{weiler2017}
{Weiler} M.,  {Hamprecht} F.~A.,   {Storath} M.,  2017, arXiv e-prints, \href
  {https://ui.adsabs.harvard.edu/abs/2017arXiv171107289W} {p. arXiv:1711.07289}

\bibitem[\protect\citeauthoryear{{Weiler}, {Geiger}, {Welling}, {Boomsma}  \&
  {Cohen}}{{Weiler} et~al.}{2018}]{weiler2018}
{Weiler} M.,  {Geiger} M.,  {Welling} M.,  {Boomsma} W.,   {Cohen} T.,  2018,
  arXiv e-prints, \href {https://ui.adsabs.harvard.edu/abs/2018arXiv180702547W}
  {p. arXiv:1807.02547}

\bibitem[\protect\citeauthoryear{{White}}{{White}}{1984}]{white1984}
{White} S.~D.~M.,  1984, \mn@doi [\apj] {10.1086/162573}, \href
  {https://ui.adsabs.harvard.edu/abs/1984ApJ...286...38W} {286, 38}

\bibitem[\protect\citeauthoryear{Wu et~al.,}{Wu et~al.}{2018}]{wu2018}
Wu C.,  et~al., 2018, \mn@doi [Monthly Notices of the Royal Astronomical
  Society] {10.1093/mnras/sty2646}, 482, 1211

\bibitem[\protect\citeauthoryear{{York} et~al.,}{{York} et~al.}{2000}]{sdss}
{York} D.~G.,  et~al., 2000, \mn@doi [\aj] {10.1086/301513}, \href
  {https://ui.adsabs.harvard.edu/abs/2000AJ....120.1579Y} {120, 1579}

\makeatother
\end{thebibliography}








\bsp	
\label{lastpage}
\end{document}